\documentclass[aps,prd,twocolumn,superscriptaddress,nofootinbib,preprintnumbers]{revtex4-1}
\usepackage{amsmath}
\usepackage{graphicx}
\usepackage{subfigure}
\usepackage{amssymb}
\usepackage{xcolor}
\usepackage{multirow}
\usepackage{cancel}
\usepackage{color}
\usepackage{ulem}
\usepackage{listings}
\usepackage{xcolor}
\usepackage{float}
\usepackage{slashed}

\usepackage{tikz}
\usetikzlibrary{arrows,shapes}
\usetikzlibrary{trees}
\usetikzlibrary{matrix,arrows} 				
\usetikzlibrary{positioning}				
\usetikzlibrary{calc,through}				
\usetikzlibrary{decorations.pathreplacing}  
\usepackage{pgffor}							

\usetikzlibrary{decorations.pathmorphing}	
\usetikzlibrary{decorations.markings}
\tikzset{
    vector/.style={decorate, decoration={snake}, draw},
	provector/.style={decorate, decoration={snake,amplitude=2.5pt}, draw},
	antivector/.style={decorate, decoration={snake,amplitude=-2.5pt}, draw},
    fermion/.style={draw=black, postaction={decorate},
        decoration={markings,mark=at position .55 with {\arrow[draw=black]{>}}}},
    fermionbar/.style={draw=black, postaction={decorate},
        decoration={markings,mark=at position .55 with {\arrow[draw=black]{<}}}},
    fermionnoarrow/.style={draw=black},
    gluon/.style={decorate, draw=black,
        decoration={coil,amplitude=4pt, segment length=5pt}},
    scalar/.style={dashed,draw=black, postaction={decorate},
        decoration={markings,mark=at position .55 with {\arrow[draw=black]{>}}}},
    scalarbar/.style={dashed,draw=black, postaction={decorate},
        decoration={markings,mark=at position .55 with {\arrow[draw=black]{<}}}},
    scalarnoarrow/.style={dashed,draw=black},
    electron/.style={draw=black, postaction={decorate},
        decoration={markings,mark=at position .55 with {\arrow[draw=black]{>}}}},
	bigvector/.style={decorate, decoration={snake,amplitude=4pt}, draw},
}

\tikzstyle{block} = [draw, rectangle, 
    minimum height=3em, minimum width=6em]

\usepackage[colorlinks,citecolor=blue]{hyperref}
\usepackage{amsmath}
\usepackage{wrapfig}

\newcommand{\be}{\begin{equation}}
\newcommand{\ee}{\end{equation}}
\newcommand{\beq}{\begin{equation}}
\newcommand{\eeq}{\end{equation}}
\newcommand{\bea}{\begin{eqnarray}}
\newcommand{\eea}{\end{eqnarray}}
\newcommand{\besp}{\begin{equation}\begin{split}}
\newcommand{\eesp}{\end{split}\end{equation}}

\newcommand{\Eq}[1]{Eq.~(\ref{#1})}

\newcommand{\Dfbd}{\mathord{\buildrel{\lower3pt\hbox{$\scriptscriptstyle\leftrightarrow$}}\over {D}_{\mu}}}
\newcommand{\ave}[1]{\left\langle #1\right\rangle}

\def\0{\textbf{0}}
\def\1{\textbf{1}}
\def\2{\textbf{2}}
\def\3{\textbf{3}}
\def\4{\textbf{4}}
\def\5{\textbf{5}}
\def\6{\textbf{6}}
\def\7{\textbf{7}}
\def\8{\textbf{8}}
\def\9{\textbf{9}}

\def\r{\textbf{r}}

\begin{document}

\title{Old Phase Remnants in First-Order Phase Transitions}

\author{Philip Lu}
\email{philiplu11@gmail.com}
\affiliation{Center for Theoretical Physics, Department of Physics and Astronomy, Seoul National University, Seoul 08826, Korea}

\author{Kiyoharu Kawana}
\email{kawana@snu.ac.kr}
\affiliation{Center for Theoretical Physics, Department of Physics and Astronomy, Seoul National University, Seoul 08826, Korea}

\author{Ke-Pan Xie}
\email{kepan.xie@unl.edu}
\affiliation{Department of Physics and Astronomy, University of Nebraska, Lincoln, NE 68588, USA}

\begin{abstract}

First-order phase transitions (FOPTs) are usually described by the nucleation and expansion of new phase bubbles in the old phase background. While the dynamics of new phase bubbles have been extensively studied, a comprehensive treatment of the shrinking old phase remnants remained undeveloped. We present a novel formalism for remnant statistics in FOPTs and perform the first calculations of their distribution. By shifting to the reverse time description, we identify the shrinking remnants with expanding old phase bubbles, allowing a quantitative evolution and determination of the population statistics. Our results not only provide essential input for cosmological FOPT-induced soliton/primordial black hole formation scenarios, but can also be readily applied to generic FOPTs.

\end{abstract}

\maketitle

\section{Introduction}

First-order phase transitions (FOPTs) are found across disciplines as diverse as biology~\cite{Lu2021.04.15.439963,Narayanan148395}, condensed matter physics~\cite{1966PhRv..141..517B,2011qpt..book.....S}, and cosmology~\cite{Sato:1980yn}. In the cosmological context, FOPTs are a natural consequence of many Beyond Standard Model theories, and could play a crucial role in generating the matter-antimatter asymmetry~\cite{Kuzmin:1985mm,Joyce:1994zt,Joyce:1994fu,Cohen:1993nk,Morrissey:2012db}, forming dark matter~\cite{Baker:2016xzo,Baker:2019ndr,Chway:2019kft,Azatov:2021ifm,Witten:1984rs,Krylov:2013qe,Huang:2017kzu,Bai:2018vik,Bai:2018dxf,Hong:2020est,Asadi:2021yml,Marfatia:2021twj} and primordial black holes (PBHs)~\cite{Crawford:1982yz,Hawking:1982ga,La:1989st,Moss:1994iq,konoplich1998formation,Konoplich:1999qq,Kodama:1982sf,Lewicki:2019gmv,Kusenko:2020pcg,Gross:2021qgx,Baker:2021nyl,Baker:2021sno,Kawana:2021tde,Marfatia:2021hcp,Huang:2022him,Liu:2021svg,Davoudiasl:2021olb,Jung:2021mku,Hashino:2021qoq,Maeso:2021xvl}, and leave detectable signals in current or near-future gravitational wave detectors~\cite{Xue:2021gyq,NANOGrav:2020bcs,Romero:2021kby,Caprini:2015zlo,Caprini:2019egz,Liang:2021bde}.

Cosmological FOPTs happen through the nucleation and growth of new true vacuum (TV) phase bubbles in the old false vacuum (FV) phase background. More attention has been focused on the calculation and estimation of the properties of TV bubbles, whose statistics are relevant for electroweak baryogenesis~\cite{Kuzmin:1985mm,Joyce:1994zt,Joyce:1994fu,Cohen:1993nk,Morrissey:2012db} and the production of gravitational waves~\cite{Hogan:1986qda,Maggiore:1999vm,Kawana:2022fum}. As a result, analytic methods were developed to estimate the TV nucleation rate, wall velocity, and bubble distribution~\cite{Callan:1977pt,Guth:1979bh,Guth:1981uk,Affleck:1980ac,Dine:1992wr,Espinosa:2010hh,Megevand:2016lpr,Ellis:2018mja,Wang:2020jrd,Bodeker:2009qy,Bodeker:2017cim,Hoche:2020ysm,Azatov:2020ufh,Gouttenoire:2021kjv,DeLuca:2021mlh}. On the other hand, FV remnants are more relevant for the mechanisms involving trapping particles in the FV to realize baryogenesis~\cite{Arakawa:2021wgz}, dark matter~\cite{Witten:1984rs,Krylov:2013qe,Huang:2017kzu,Bai:2018vik,Bai:2018dxf,Hong:2020est,Asadi:2021yml,Marfatia:2021twj} and PBHs~\cite{Baker:2021nyl,Kawana:2021tde,Baker:2021sno,Marfatia:2021hcp,Huang:2022him}.
Lacking an equivalent detailed description of the FV remnants, previous studies had to either use naive estimations of the average remnant size and density~\cite{Krylov:2013qe,Hong:2020est}, or take those observables as free parameters~\cite{Baker:2021nyl,Baker:2021sno}.

\begin{figure}
\begin{center}
\includegraphics[scale=0.42]{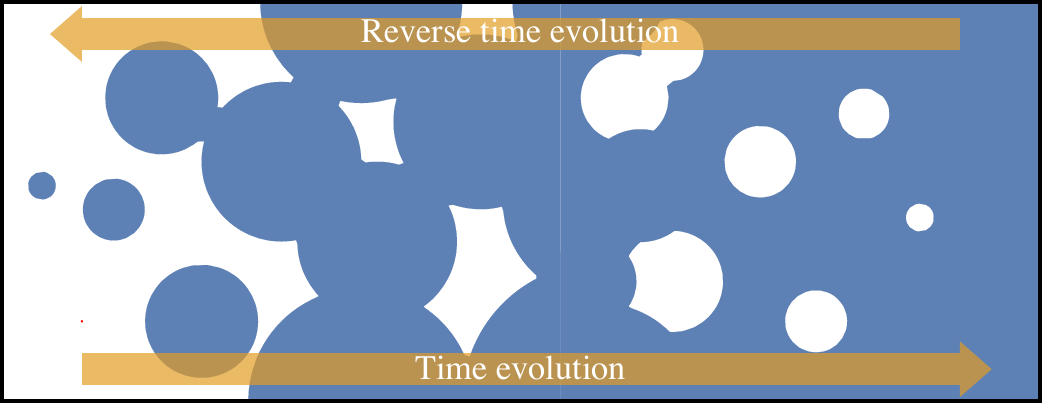}
\caption{Progression of the FOPT: with the flow of time from left to right, the initial FV (white) space is nucleated with TV (blue) bubbles, which populate and surround the shrinking FV bubbles. In the reverse time description, FV bubbles nucleate and grow in a TV dominated space, eventually resulting in shrinking TV bubbles.}
\label{fig:FOPT}
\end{center}
\end{figure}

In the existing framework, the FV phase acquires a decay probability to the energetically favorable TV phase below the critical temperature.  The vacuum pressure causes the TV bubbles to expand, filling up the space and leaving shrinking pockets of disconnected FV remnants, see Fig.~\ref{fig:FOPT}. We extend the TV bubble nucleation formalism to include FV bubble nucleation by considering the phase transition in reverse. From this reverse time description, the centers of the collapsing remnants with time flowing forward can be viewed as the nucleation sites of FV bubbles with time flowing backwards. Thus, the methods used for TV bubbles can be adapted for the dynamics of FV bubbles. We perform the first calculation of the properties and evolution of the FV bubbles based on the {\it reverse} time description of the FOPT. Key to the validity of this method is the projection interpretation, in which we perform our calculations using the extrapolated evolution of the FOPT as a mathematical tool. Although we present our results within a cosmological context, they can be easily adapted to general FOPTs in other fields.

In this paper, we develop a method for calculating remnant statistics from first principles. In Section~\ref{sec:bubble}, we survey the formalism for TV bubble nucleation and connect it to FV bubbles. In Section~\ref{sec:wallcone}, we derive a general expression for the FV nucleation rate in one, two, and three dimensions. In Section~\ref{sec:dist}, the FV bubble distribution is explicitly calculated in the exponential approximation and applied to PBH formation. In Section~\ref{sec:discussion}, we elaborate on the projection interpretation and summarize our work. Details of the angular integration are found in Appendix~\ref{sec:appendix}.

\begin{figure*}
\begin{center}
\includegraphics[scale=0.35]{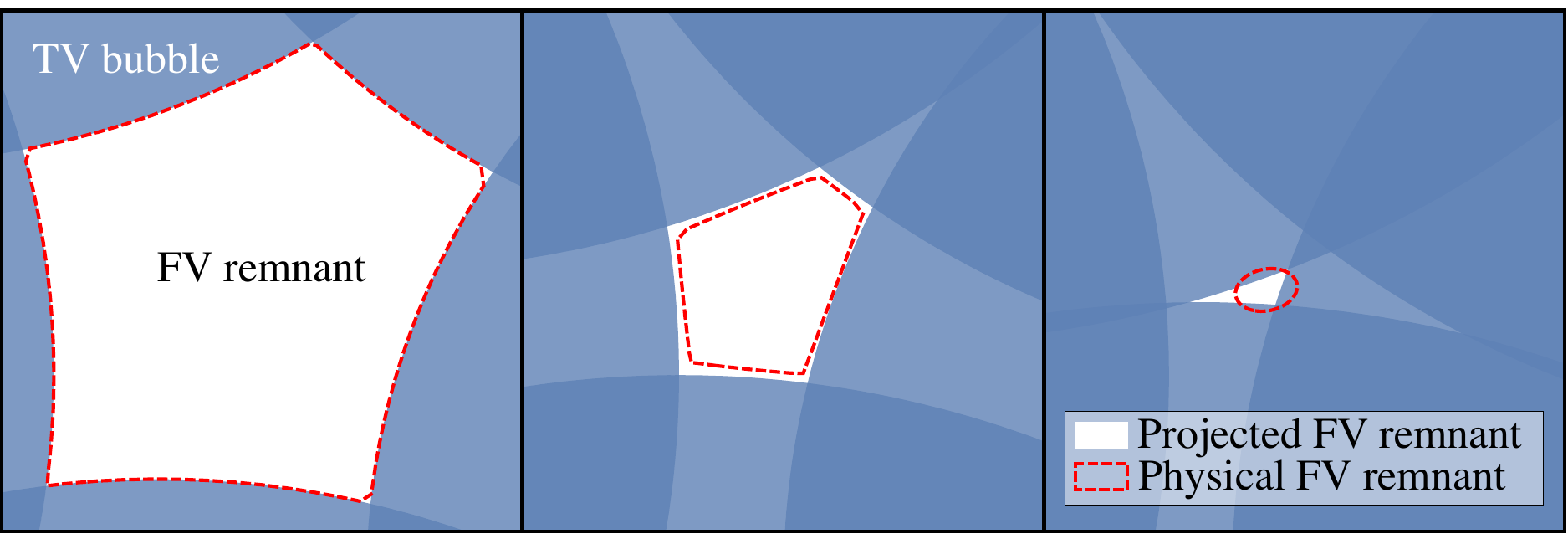}
\caption{Illustration of the projection interpretation. Due to surface tension, the shape of a physical FV remnant will eventually tend to be spherical, as shown in red dashed line. In contrast, the projected FV remnant shrinks as if the TV bubbles do not interact with each other at all, and eventually becomes a triangle/tetrahedron in two/three spatial dimension case, as shown by the white region.}
\label{fig:projection}
\end{center}
\end{figure*}

\section{Bubble Formation}\label{sec:bubble}

\subsection{True Vacuum Bubbles}

We review the existing TV nucleation formalism for cosmological phase transitions. We assume thin walls and constant velocity $v_w$ throughout, which is a good approximation for a range of moderately strong FOPTs~\cite{Megevand:2016lpr}. Consider a Universe initially in an FV phase at high temperatures. Below the critical temperature $T_c$, the TV phase becomes energetically favorable, giving a non-zero probability for the FV space to tunnel to the TV. The TV nucleation rate per unit volume and unit time is~\cite{Affleck:1980ac}
\begin{equation}\label{Gamma}
\Gamma(t) = A(t) e^{-S(t)},
\end{equation}
where $S(t)$ is the smaller of the two instanton bounce actions $S_3/T$~\cite{Linde:1981zj} and $S_4$~\cite{Coleman:1977py}.

Assuming the bubbles grow spherically outwards, the radius of a bubble nucleated at time $t'$ is
\be\label{Rt}
R(t,t')=v_w\int_{t'}^tdt''\frac{a(t)}{a(t'')}~,
\ee
where $a(t)$ is the scale factor of the FLRW metric. The fraction of space in the FV is ~\cite{Guth:1979bh,Guth:1981uk}
\begin{equation}\label{eq:fplus}
f_{\rm fv}(t) = e^{-I(t)}~,
\end{equation}
where
\begin{equation}
\label{eq:iplus}
I(t) = \int_{t_c}^{t} dt^\prime \Gamma(t^\prime) \frac{a^3(t^\prime)}{a^3(t)}\frac{4\pi}{3}R^3(t,t')~,
\end{equation}
with $t_c$ being the cosmic time corresponding to $T_c$.

The distribution of TV bubbles at time $t$ with size $R$ must equal the average nucleation rate at a time $t'$ at which $R(t,t')=R$ (when the scale factor is negligible, $t'=t-R/v_w$),
\begin{equation}\label{eq:dndr}
\frac{dn_{\rm tv}}{dR}(t) =\frac{1}{v_w}f_{\rm fv}(t')\Gamma(t')\frac{a^4(t')}{a^4(t)}~,
\end{equation}
where the factor of $f_{\rm fv}$ restricts the nucleation to the false vacuum. Integrating this equation, the total bubble number density is given by
\be\label{n(t)}
n_{\rm tv}(t)=\int_{t_c}^tdt' f_{\rm fv}(t')\Gamma(t')\frac{a^3(t')}{a^3(t)}~,
\ee
where we have used Eq.~(\ref{Rt}). 

\subsection{False Vacuum Bubbles}

In the forward evolution of the FOPT, spherical TV bubbles percolate when the FV volume fraction drops below $f_{\rm fv} = 0.71$~\cite{rintoul1997precise}, forming an infinite connected cluster. As $f_{\rm fv}$ decreases, FV regions are separated into shrinking remnants, which eventually tend to be spherical due to surface tension. When the process is considered in reverse, roughly spherical FV bubbles nucleate in a TV background, forming an infinite connected cluster around $f_{\rm fv}\approx 1-0.71=0.29$. This is the reverse time description of the FOPT, in which we identify shrinking remnants forward in time with nucleating bubbles backward in time.

We develop a formalism for FV bubbles in this reverse time description and calculate a FV bubble nucleation rate. To do so, the nucleation point of a FV bubble can be identified as the projected center of a collapsing remnant, with the nucleation rate equal to the collapse rate. The idea is illustrated in Fig.~\ref{fig:projection}, where we have plotted and compared the shapes of two types of FV remnants. The physical FV remnant, which really exists in a FOPT, as shown in red dashed line, eventually changes its shape to be more and more spherical due to surface tension. The projected FV remnant, which is an imaginary object assuming the TV bubbles do not interact with each other, eventually shrinks into a triangle/tetrahedron shape in the two/three spatial dimensions case when the size of the remnant is much smaller than the radii of the enveloping TV bubbles. In the projection interpretation, we analytically calculate the evolution of the projected FV remnants from the percolation time to the final collapse, in order to describe the state of the physical FV remnants at the percolation time. This is the core idea of this article.

The projection interpretation, in which we extrapolate the wall trajectories to the point of collapse, suggests a counterpart to the reverse time description. In the reverse time description, the FV bubbles nucleated at point $C$ and time $t'>t$ grow to a size $R_r(t,t')$ at time $t$. In the alternative forward time description, the FV remnants of size $R_r(t,t')$ at time $t$ are those whose walls are projected to collapse at point $C$ and time $t'>t$. These two viewpoints are complementary and fundamentally equivalent.

The probability of remnant collapse per unit volume per unit time, $d^2P_r/dVdt$, at a point $C$ can be found by integrating along the past {\it wall cone} (similar to the light cone but with wall velocity $v_w$) and finding the four points of TV nucleation, so that the four resulting walls would meet at point $C$ and time $t$. We order the TV walls by their proximity in time to $t$, with wall 1 being the most recently nucleated wall and thereby avoid unnecessary combinatoric factors. The nucleation point of wall 1 is only constrained to lie on the past wall cone of point $C$, but subsequent walls have to obey angular restrictions to form a closed remnant. As the nucleations lie on the wall cone surface, assuming constant velocity, the walls are always equidistant from the center point. Due to this symmetry, the radial/temporal and angular factors are independent. We first introduce the formalism in one and two spatial dimensions before developing the more complicated three spatial dimensions case.

\section{Wall Cone Integration}\label{sec:wallcone}

\subsection{One Spatial Dimension}

\begin{figure}
\begin{center}
\includegraphics[scale=0.4]{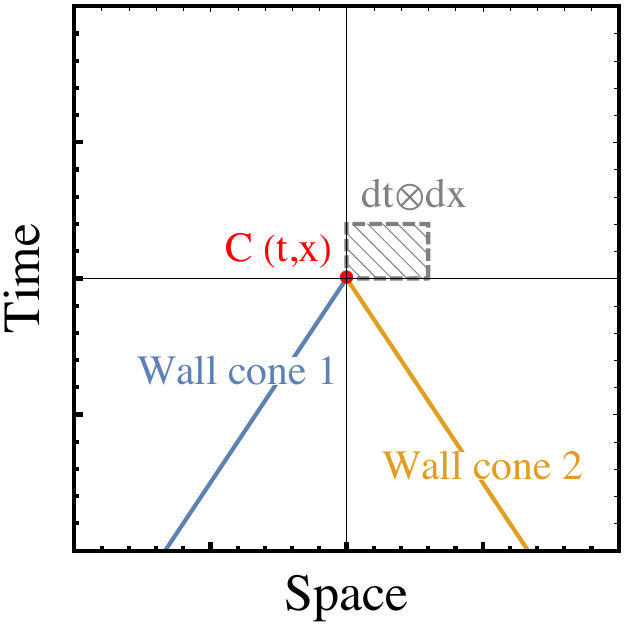}
\caption{Sketch of two walls collapsing at point $C$ in the $1+1$ spacetime case.}
\label{fig:collapse_1d}
\end{center}
\end{figure}

As an illustration of the main idea, let us first consider the one spatial dimension case shown in Fig.~\ref{fig:collapse_1d}, where the TV bubbles that nucleate on surfaces of wall cone 1 (blue line) and 2 (yellow line) collapse at point $C$ (red dot). The key point here is that {\it all} TV bubbles that can reach $C$ from the $-\hat x$ ($+\hat x$) direction must nucleate on the surface of wall cone 1 (2). Therefore, integrating along the wall cones provides the probabilities of walls collapsing at $C$.

The physical event  ``two walls collapse at the same point $C(t,x)$'' can be decomposed into three independent subevents. The first one is wall 1 from the $-\hat x$ direction approaching at the space region $[x,x+dx]$ at time $t$. The corresponding probability can be derived by integrating along the wall cone,
\be
dP_1=dx\int_{t_c}^tdt_1\Gamma(t_1),
\ee
where the scale factor is omitted in this subsection for simplicity. Similarly, the second subevent is wall 2 from the $+\hat x$ direction approaching at point $x$ but within the time region $[t,t+dt]$, and the probability is
\be
dP_2=v_wdt\int_{t_c}^tdt_2\Gamma(t_2).
\ee

There is, however, an important third subevent, which is {\it no TV bubble reaches space point $x$ before $t$}, or equivalently, no TV bubble nucleates in the region below wall cones 1 and 2 (or say, inside the wall cone) in Fig.~\ref{fig:collapse_1d}. This is actually the probability of $C$ lying in the FV region, i.e. $f_{\rm fv}(t)$ given in \Eq{eq:fplus}. Combining the probabilities of the three subevents, we eventually reach the collapsing probability density
\be
\frac{d^2P}{dxdt}=v_wf_{\rm fv}(t)\int_{t_c}^tdt_1\Gamma(t_1)\int_{t_c}^{t_1^{}}dt_2\Gamma(t_2).
\ee
This can be defined as the nucleation probability of the FV bubbles, i.e. $\Gamma_r^{\rm 1d}(t)$.

\subsection{Two Spatial Dimensions}

For simplicity, we omit scale factors in this exposition and restore them in the full three-dimensional expression Eq.~\eqref{eq:3dgammar}. In its final stage, the infinitesimal shrinking FV remnant is a triangle collapsing towards its incenter surrounded by three TV walls. The collapse probability per unit area per unit time, $d^2P_{r}/dAdt$ can be found by integrating along the past wall cone of the collapse point $C$. The radial and angular integrations can be separated, and the radial part is
\begin{multline}\label{eq:2drad}
v_w f_{\rm fv}(t)\int_{t_c}^tdt_1v_w(t-t_1)\Gamma(t_1) \int_{t_c}^{t_1}dt_2 v_w(t-t_2)\Gamma(t_2)\\
\times \int_{t_c}^{t_2}dt_3v_w(t-t_3)\Gamma(t_3)~,
\end{multline}
with the integration times ordered as $t_c<t_3<t_2<t_1<t$, corresponding to the nucleation time of each successive wall. 
The factors of $\Gamma$ gives the TV nucleation rate of each wall, and the factor of the FV fraction, $f_{\rm fv}$, is required since only FV points are eligible to collapse to TV. Since we are integrating along the past wall cone, the integration space of the walls is in the FV and no additional factors of $f_{\rm fv}$ show up inside the integrals. Otherwise, there would need to be a TV nucleation inside the within the past wall cone which would have already spread the TV point $C$ before time $t$. This would contradict the overall factor of $f_{\rm fv}$ imposed on point $C$.

For the angular factor, we denote the normal vector of wall $i$ by $\hat\r_i$, i.e. the unit vector pointing from the TV nucleation point towards $C$, the incenter of the triangle. For the three walls to form a closed triangle, $-\hat\r_3$ must lie within the angular range bounded by $\hat\r_1$ and $\hat\r_2$, as illustrated in the left panel of Fig.~\ref{fig:collapse}. Integration over this restricted angular parameter space yields an additional factor of $2\pi^3$ (see Appendix~\ref{sec:appendix} for a derivation).

In the reverse time description, the FV bubble nucleation rate is defined as the remnant collapse probability per unit area per unit time, $\Gamma_r^{\rm 2d}(t)=d^2P_r/dAdt$. Combining Eq.~\eqref{eq:2drad} and the angular contribution, the FV bubble nucleation rate in two dimensions is
\begin{multline}\label{eq:2ddpdvdr}
\Gamma_r^{\rm 2d}(t)=2\pi^3v_w^4f_{\rm fv}(t)\int_{t_c}^tdt_1(t-t_1)\Gamma(t_1)\times\\
\int_{t_c}^{t_1}dt_2(t-t_2)\Gamma(t_2)
\int_{t_c}^{t_2}dt_3(t-t_3)\Gamma(t_3)~.
\end{multline}

\subsection{Three Spatial Dimensions}

\begin{figure}
\begin{center}
\includegraphics[scale=0.35]{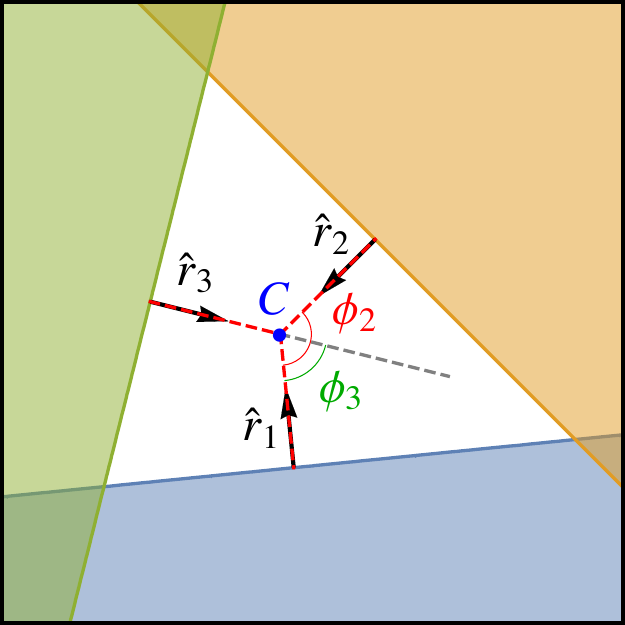}\qquad
\includegraphics[scale=0.32]{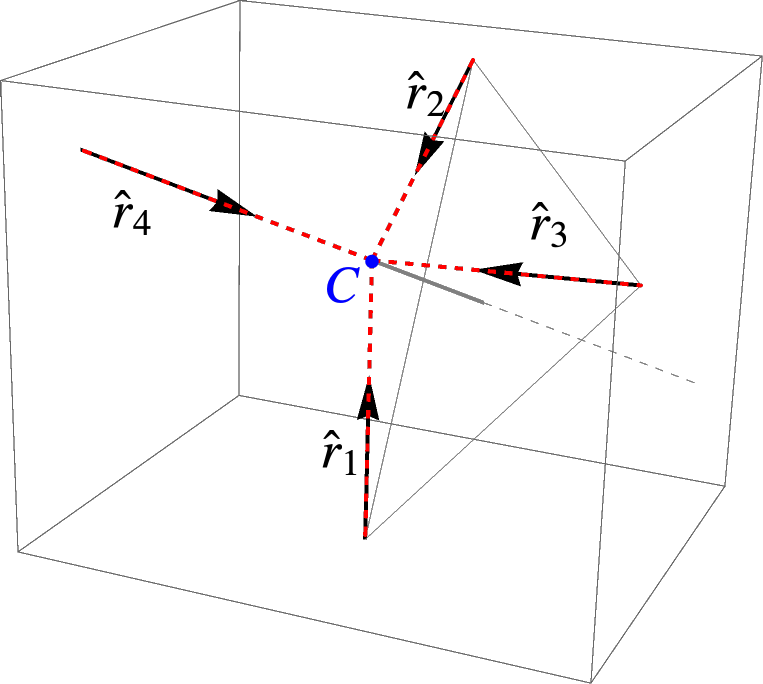}
\caption{Geometry of a collapsing FV remnant. \textbf{Left:} In 2-dimensional space, to form a closed triangle, $-\hat\r_3$ must lie in the angular range bounded by $\hat\r_1$ and $\hat\r_2$. \textbf{Right:} In 3-dimensional space, to form a closed tetrahedron, $-\hat\r_4$ must lie in the solid angle delimited by $\hat\r_1$, $\hat\r_2$ and $\hat\r_3$ (the four walls are omitted for clarity).}
\label{fig:collapse}
\end{center}
\end{figure}

Next we build on our two dimensional results and apply the method to three dimensions. Integrating over the past wall cone, the radial factor is similar, but the angular factor is more complicated, as the three-dimensional remnant is now tetrahedral with four collapsing walls. Ordering the walls temporally and labeling the normal vectors as before, the condition that the four walls form a tetrahedron is that $-\hat\r_4$ should lie in the solid angle delimited by $\hat\r_1$, $\hat\r_2$ and $\hat\r_3$, as illustrated in the right panel of Fig.~\ref{fig:collapse}. This leads to an overall angular factor of $32\pi^4$ (see Appendix~\ref{sec:appendix} for a derivation).
The FV bubble nucleation rate at time $t$ is then
\begin{widetext}
\begin{multline}\label{eq:3dgammar}
\Gamma_r(t)\equiv\frac{d^2P_r}{dVdt}=32\pi^4v_w^9f_{\rm fv}(t) \int_{t_c}^tdt_1\left(\int_{t_1}^tdt_1'\frac{a(t_1)}{a(t_1')}\right)^2\Gamma(t_1)\frac{a(t_1)}{a(t)}\int_{t_c}^{t_1}dt_2\left(\int_{t_2}^tdt_2'\frac{a(t_2)}{a(t_2')}\right)^2\Gamma(t_2)\frac{a(t_2)}{a(t)} \\ \times 
\int_{t_c}^{t_2}dt_3\left(\int_{t_3}^tdt_3'\frac{a(t_3)}{a(t_3')}\right)^2\Gamma(t_3)\frac{a(t_3)}{a(t)}\int_{t_c}^{t_3}dt_4\left(\int_{t_4}^tdt'_4\frac{a(t_4)}{a(t'_4)}\right)^2\Gamma(t_4)\frac{a(t_4)}{a(t)}~,
\end{multline}
\end{widetext}
where the integrals over the scale factors come from the integration element of the radial direction in spherical coordinates. Eq.\eqref{eq:3dgammar} is a general formula for the FV nucleation rate for arbitrary TV nucleation rate $\Gamma$ and expansion history. The integral simplifies to
\be
\left(\int_{t_i}^tdt_i'\frac{a(t_i)}{a(t_i')}\right)^2\to (t-t_i)^2~,
\ee
when the scale factors can be taken as constant.\\

\section{False Vacuum Bubble Distribution}
\label{sec:dist}
In the reverse time description, we can use the FV bubble nucleation rate, \Eq{eq:3dgammar}, to find the FV bubble distribution. Analogously to the TV bubble case, a FV bubble which nucleates at $t'>t$ has a radius
\be\label{Rrt}
R_r(t,t')=v_w\int_t^{t'}dt''\frac{a(t)}{a(t'')}~,
\ee
at time $t$. 

The FV bubble size distribution is then
\begin{equation}
\frac{n_{\rm fv}}{dR_r}(t)=\frac{1}{v_w}(1-f_{\rm fv}(t'))\Gamma_r(t')\frac{a^4(t')}{a^4(t)}~,\label{eq:remnantnumden}
\end{equation}
where $t'$ is resolved using \Eq{Rrt}, reducing to $t'=t+R_r/v_w$ in the limit of constant scale factor. Integrating \Eq{eq:remnantnumden} over $R_r$ yields the overall FV bubble number density
\be
n_{\rm fv}(t)=\int_t^{t_e}dt'(1-f_{\rm fv}(t'))\Gamma_r(t')\frac{a^3(t')}{a^3(t)}~,
\ee
where $t_e$ is the ending time of the FOPT, which can be effectively taken as $+\infty$ and $f_{\rm fv}(t_e)=0$.\\

\subsection{Exponential Nucleation}

So far, our results are rather general and apply to any FOPT scenario as long as the TV nucleation rate $\Gamma(t)$ is available. We now evaluate our results in the exponential nucleation rate approximation,
\begin{equation}
\label{eq:expnuc}
\Gamma(t) = \Gamma_\ast e^{\beta(t-t_\ast)}~,
\end{equation}
expanded around an arbitrary time $t_*\in(t_c,t_e)$, where $\beta=-dS(t)/dt|_{t_*}$ can be treated as the inverse of the FOPT duration. This exponential approximation is accurate if the FOPT proceeds rapidly compared to the Hubble time scale, i.e. $\beta/H(t_*)\gg1$~\cite{Megevand:2016lpr,Turner:1992tz}.\footnote{For a long duration FOPT, the Gaussian nucleation rate approximation is more suitable~\cite{Megevand:2016lpr}.} Hence, the scale factor $a(t)$ is approximately constant over the transition and can be neglected. This treatment will be adopted for the rest of this article.

The exponential approximation fits numerical simulations the best when $t_\ast$ is chosen to be the time at which the bubble statistics are computed. For the remnant distribution, we choose $t_*$ to be the FV bubble percolation time, which we approximate as $f_{\rm fv}(t_\ast) = 0.29$ and $I_\ast=-\ln(0.29)= 1.238$.

We explicitly solve for the FV filling fraction, Eq.~\eqref{eq:fplus} with 
\begin{equation}\label{eq:expI}
I(t)=\frac{8\pi v_w^3 \Gamma_\ast}{\beta^4}e^{\beta(t-t_\ast)} \equiv I_\ast e^{\beta(t-t_\ast)}~,
\end{equation}
using $\beta(t_*-t_c)\gg1$. Integrating Eq.~\eqref{eq:3dgammar}, the FV bubble nucleation rate is
\be\label{eq:3dexpgammar}
\Gamma_r(t)\approx\frac{I_*^4\beta^4}{192v_w^3}e^{4\beta(t-t_*)}e^{-I(t)}~.
\ee
Using Eq.~(\ref{eq:remnantnumden}), the FV bubble distribution in the constant velocity, constant scale factor exponential approximation is
\begin{multline}\label{eq:3ddndr}
\frac{dn_{\rm fv}}{dR_r}(t_*)\approx\frac{I_*^4\beta^4}{192v_w^3}e^{4\beta R_r/v_w} e^{-I_*e^{\beta R_r/v_w}}\\
\times\left(1-e^{-I_*e^{\beta R_r/v_w}}\right)~,
\end{multline}
where the factor of $1-f_{\rm fv}(t)$ comes from FV bubbles nucleating only within the TV.

\subsection{Normalization}

We determine the approximate shape of the FV bubbles by normalizing the total volume contained in the FV bubbles to the filling fraction $f_{\rm fv}$ at the remnant percolation time $t_*$. The initially tetrahedral FV bubble is expected to become more rounded as additional TV walls partially cover the bubble, mimicking the effects of tension neglected in this treatment. We expect FV bubbles at the percolation threshold to be roughly spherical with some abnormality, as depicted in Fig.~\ref{fig:FOPT}. We therefore parameterize the volume formula as $A(4\pi/3) R^3$. Here $R$ is the minimum distance between the central point $C$ and the remnant walls so $A=1$ implies a spherical volume and $(A-1)$ measures the departure from sphericality. Normalizing to the FV fraction,
\be
\int dR_rA \frac{4\pi R_r^3}{3}\frac{dn_{\rm fv}}{dR_r}(t_*)=0.29~,
\ee
yields $A = 1.15$. This suggests that the FV bubbles/remnants are somewhat spherical. From another point of view, the factor $A=1.15$ is the required normalization for the formalism to be self-consistent. Although overlap between adjacent FV bubbles is ignored here, the volume $V\approx 1.15(4\pi/3)R^3$ is the average volume that effectively ``belongs'' to each collapsing remnant of size $R$ at the percolation time.

Since the value of the FV percolation chosen here, $f_{\rm fv}=0.29$, is strictly valid only for spherical bubbles of equal size, the exact percolation threshold would have to be determined by numerical simulation. The volume factor $A$ is reasonably close to 1, so we do not expect the true percolation threshold to significantly deviate from the approximate value used.

\subsection{Primordial Black Holes}

As a concrete example, we apply our method to the Fermi-ball/PBH formation scenario proposed in Refs.~\cite{Hong:2020est,Kawana:2021tde} and subsequently studied in Refs.~\cite{Marfatia:2021twj,Marfatia:2021hcp,Huang:2022him}. During the FOPT, an asymmetric population of dark fermions $\chi$-$\overline{\chi}$ is trapped between the expanding TV bubble walls into the collapsing FV remnants due to a large mass differential for $\chi$ in the two phases. As the remnants shrink, the fermions and anti-fermions annihilate, leaving only the asymmetrical portion supported by degeneracy pressure. The total number of $\chi$ fermions trapped in a remnant with size $R_*$ at the FV percolation time $t_*$ is~\cite{Hong:2020est,Kawana:2021tde}
\begin{equation} \label{eq:qfb}
Q_{\rm FB} = \frac{\eta_\chi s(t_*)}{f_{\rm fv}(t_*)}A\frac{4\pi R_*^3}{3}~,
\end{equation}
where $\eta_\chi=(n_\chi-n_{\bar\chi})/s$ is the $\chi$-asymmetry with $s$ the entropy density.

\begin{figure}
\begin{center}
\includegraphics[scale=0.45]{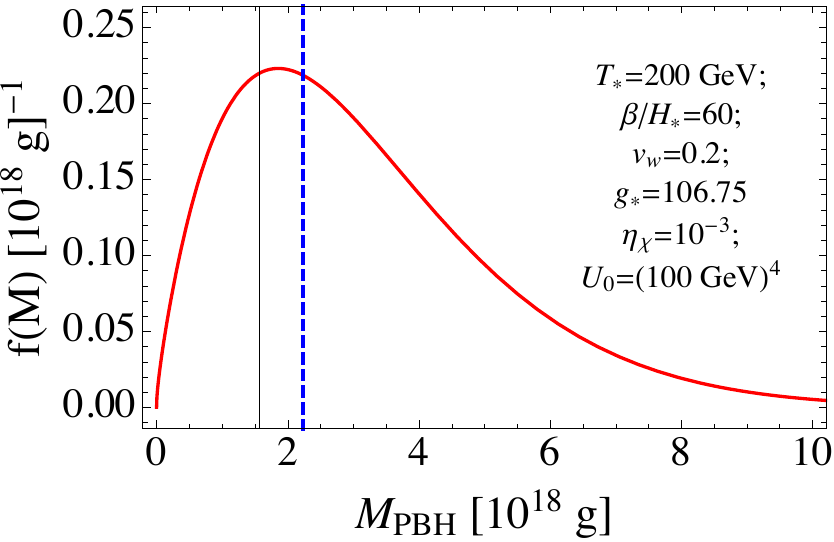}
\caption{A benchmark $f(M_{\rm PBH})$ distribution (red curve) for the Fermi-ball/PBH formation scenario of Ref.~\cite{Kawana:2021tde} calculated with an exponential nucleation rate at constant scale factor, where $f$ is the PBH mass distribution probability function with the FOPT parameters shown in the figure. We compare $\ave{M_{\rm PBH}}$ (black solid line) from this work and the estimate of $M_{\rm PBH}$ (blue dashed line) from Ref.~\cite{Kawana:2021tde}.}
\label{fig:MdndM}
\end{center}
\end{figure}

Since the Fermi-ball/PBH mass $M_{\rm PBH}\propto Q_{\rm FB}$~\cite{Hong:2020est}, the distribution of $R_*$ is key to deriving the Fermi-ball/PBH mass profile. Lacking methods to compute the $R_*$ distribution, Refs.~\cite{Hong:2020est,Kawana:2021tde} estimated the average size to be
\be\label{estimatedR}
R_*\sim\left(\frac{3v_w}{4\pi A\Gamma_*}\right)^{1/4}=1.43\frac{v_w}{\beta}~,
\ee
resulting in a monochromatic Fermi-ball/PBH mass distribution. With the technique developed in this work, we compute the average $R_*$ using \Eq{eq:3ddndr},
\be
\ave{R}_*=\int dR_rR_r\frac{dn_{\rm fv}}{dR_r}(t_*)=1.12\frac{v_w}{\beta}~.
\ee
Although the difference is minor, a continuous $R_*$ distribution results in an extended Fermi-ball/PBH mass profile, which greatly impacts experimental constraints~\cite{Carr:2017jsz,Green:2020jor}. In Fig.~\ref{fig:MdndM}, we display an example PBH distribution at present time which comprises all of dark matter within the PBH mass window, $10^{17}~{\rm g} < M <10^{23} ~{\rm g}$ ~\cite{Carr:2020gox}.
\\

\section{Discussion}
\label{sec:discussion}
In our derivation of the FV bubble nucleation rate Eq.~\eqref{eq:3dgammar}, we made a few simplifying assumptions. First, the walls were assumed to be infinitesimally thin and the wall velocity constant. TV bubble mergers can alter the effective location of the TV wall nucleation. Furthermore, surface tension tends to shape the collapsing remnants to be more spherical, whereas the collapsing remnant is treated as tetrahedral. All of these effects are exacerbated near the end of the phase transition when TV bubbles inevitably merge, surface tension becomes more prominent as the surface area to volume ratio increases, and particle trapping may stop or slow the collapse.

We resolve these issues by interpreting the formalism for computing $\Gamma_r(t)$ as a projected description of the FOPT rather than a physical description. In other words, beyond the remnant percolation time $t_*$ at which we evaluate Eq.~\eqref{eq:3ddndr}, the future evolution of the FOPT is irrelevant to the remnant statistics evaluated at $t_*$. Causally, the remnant size distribution and number density at $t_*$ cannot be affected by events at later times $t>t_*$. Thus, the remnant statistics at the remnant percolation time depend only on the past history of the FOPT, during which these four assumptions are only mildly violated. The observables will be the same whether, in the later stages of the transition, our idealized collapse scenario or a more physical scenario with surface tension effects is applied. Hence, the projection interpretation is that our method traces the walls of the collapsing remnants forward beyond time $t_*$ to find the collapse point, in order to then trace the collapse backwards in the reverse time description and infer the size of the remnant at time $t_*$. 

We offer an analogy: the shadow of a falling apple can be used to infer its instantaneous position. Whether or not the apple eventually lands directly on its shadow, is perturbed by a gust of wind, or drops on an unwitting head, is immaterial to the determination of its instantaneous position. Likewise, whether the remnant eventually collapses spherically or is stopped by degeneracy pressure is irrelevant to its size distribution at remnant percolation. Therefore, in the derivation of the remnant distribution, our tetrahedral collapse model is more appropriate than a physically realistic spherical collapse model because the shadow (or projection) of the collapsing walls is tetrahedral and not spherical.

In summary, we have performed the first calculation of the FV remnant distribution and evolution in FOPTs. By identifying the center of a collapsing remnant as a FV bubble nucleation in reverse, the established TV bubble nucleation formalism can be adapted to derive remnant statistics. Our results provide a more sophisticated treatment of FOPT remnants than previously available, and are directly applicable to many new physics mechanisms involving trapped particles~\cite{Arakawa:2021wgz,Witten:1984rs,Krylov:2013qe,Huang:2017kzu,Bai:2018vik,Bai:2018dxf,Hong:2020est,Asadi:2021yml,Marfatia:2021twj,Baker:2021nyl,Kawana:2021tde,Baker:2021sno,Marfatia:2021hcp,Huang:2022him}. The novel formalism developed in this paper can be readily generalized and applied outside the cosmological context.\\

\acknowledgments

We thank Sunghoon Jung, Hyung Do Kim, Taehun Kim, and Howard Chen for useful discussions. K.P.X. is supported by the University of Nebraska-Lincoln. The work of K.K. and P.L. is supported by Grant Korea NRF-2019R1C1CC1010050, 2019R1A6A1A10073437.  

\appendix

\section{Angular Contribution}
\label{sec:appendix}
For the two-dimensional angular factor, we denote the normal vector of wall $i$ by $\hat\r_i$, i.e. the unit vector pointing from the TV nucleation point towards $C$. For the three walls to form a closed triangle, $-\hat\r_3$ must lie in the angular range bounded by $\hat\r_1$ and $\hat\r_2$, as illustrated in the left panel of Fig. 2 of the main text. To integrate over all the allowed TV bubble configurations, we first choose $\hat\r_1$ along the $\hat x$ axis and parameterize the other two normal vectors as
\be
\hat\r_2=(\cos\phi_2,\sin\phi_2),\quad \hat\r_3=(-\cos\phi_3,-\sin\phi_3)~.
\ee
Then for $0<\phi_2<\pi$, the closure of the FV triangle requires $0<\phi_3<\phi_2$; while the case $\pi<\phi_2<2\pi$ is just a reflection of the case $0<\phi_2<\pi$ across the $\hat x$ axis. Therefore, the angular integral reads
\be \label{eq:2dang}
2\pi\int_0^\pi 2d\phi_2\int_0^{\phi_2}d\phi_3=2\pi^3~,
\ee
where the first ``$2\pi$'' factor represents the integral over the arbitrary $\hat\r_1$ angle, while the ``2'' factor in $\phi_2$ integral accounts for the $\pi<\phi_2<2\pi$ region.

Similarly, in three dimensions, with the collapse point $C$ at the origin, we set the $\hat z$ axis in the direction of $\hat\r_1$ and $\hat y$ axis along the $\hat\r_1\times\hat\r_2$ direction, defining our spherical coordinate system. The normal vectors can then be parameterized as 
\begin{align}
\begin{split}
&\hat\r_2=(\sin\theta_2,0,\cos\theta_2),\quad \\
&\hat\r_3=(\sin\theta_3\cos\phi_3,\sin\theta_3\sin\phi_3,\cos\theta_3),\quad \\
&\hat\r_4=(-\sin\theta_4\cos\phi_4,-\sin\theta_4\sin\phi_4,-\cos\theta_4)~.
\end{split}
\end{align}
When $0<\phi_3<\pi$, the closure condition requires $0<\phi_4<\phi_3$ and $0<\theta_4<\bar\theta(\phi_4)$, where
\begin{multline}\label{theta4}
\bar\theta(\phi_4)={\rm arccot}\left(\cot\theta_2\cos\phi_4\right.\\
\left.+\frac{\cot\theta_3-\cot\theta_2\cos\phi_3}{\sin\phi_3}\sin\phi_4\right)~
\end{multline}
is determined by the intersection of plane $C\hat\r_4\hat z$ [$(\sin\phi_4)x-(\cos\phi_4)y=0$] and plane $C\hat\r_2\hat\r_3$ [$(\cot\theta_2) x+(\sin\phi_3^{})^{-1}(\cot\theta_3^{}-\cot\theta_2^{}\cos\phi_3^{})y-z=0$]. Here the range of the arccotangent function is limited to $(0,\pi)$. As the case $\pi<\phi_3<2\pi$ is just a reflection of the case $0<\phi_3<\pi$ over the $C\hat x\hat z$ plane. 
Therefore, the angular integral reads
\begin{multline}\label{3d_angle}
4\pi\int_0^\pi d\theta_2\int_0^\pi d\theta_3\int_0^{\bar\theta(\phi_4)}d\theta_4
\int_0^{2\pi}d\phi_2
\int_0^\pi 2d\phi_3 \\
\times \int_0^{\phi_3}d\phi_4\sin\theta_2\sin\theta_3\sin\theta_4=32\pi^4~,
\end{multline}
where the first ``$4\pi$'' factor represents the integral over the arbitrary $\hat\r_1$, while the factor of ``2'' in the $\phi_3$ integral accounts for the $\pi<\phi_3<2\pi$ region. 

Alternatively, we can derive Eq.~\eqref{3d_angle} in a more intuitive manner. Requiring  $-\hat\r_4$ to be in the solid angle established by $\hat\r_1$,$\hat\r_2$,and $\hat\r_3$ results in the two conditions
\begin{equation}
\label{eq:phicond}
    \pi \leqslant \phi_4 \leqslant \pi+\phi_3~,
\end{equation}
and
\begin{equation}
\label{eq:thetacond}
    \max(\theta_2,\theta_3,\theta_4) \geqslant \pi-\min(\theta_2,\theta_3,\theta_4) ~.
\end{equation}
The $\phi$ condition can be understood as the requirement that walls 2, 3 and 4 form a triangle in the plane of wall 1, and so is analogous to the two-dimensional case Eq.~\eqref{eq:2dang}. The $\theta$ condition can be interpreted as the need to form a closed tetrahedron in the $\hat{z}$ direction. To understand this condition, first sort the angles $\theta_2,\theta_3,\theta_4\xrightarrow{} \theta_l\leq\theta_m\leq\theta_h$. Take the limiting case where wall $h$ is arranged opposite of wall $l$, so that the corresponding azimuthal angles satisfy $\phi_h = \pi+\phi_l$. We see that the closure requirement is satisfied if $\pi-\theta_l \leq \theta_h \leq \pi$, where at the lower bound wall $h$ lies directly opposite of wall $l$ and at the upper bound, wall $h$ is directly opposite wall $1$. Therefore, the $\phi$ contribution to the three-dimension probability integral is
\begin{equation}
\label{eq:phicont}
    \int_{0}^{2\pi} d\phi_1 \int_{0}^{2\pi} d\phi_2 \int_{0}^{\pi} 2d\phi_3 \int_{\pi}^{\pi+\phi_3} d\phi_4 = 4\pi^4~,
\end{equation}
and the $\theta$ contribution is
\begin{multline}\label{eq:thetacont}
\int_{0}^{\pi}d\theta_1 \sin\theta_1~ 3! \int_{0}^{\pi} d\theta_l \sin\theta_l \int_{\max(\pi-\theta_l,\theta_l)}^{\pi} d\theta_h \sin\theta_h \\  \times \int_{\theta_l}^{\theta_h} d\theta_m \sin\theta_m = 8~.
\end{multline}
The combinatoric factor of $3!$ comes from the sorting of $\theta_2$,$\theta_3$, and $\theta_4$ into the low, medium, and high angles. As in Eq.~\eqref{3d_angle}, the combined angular contribution is $32\pi^4$.

\bibliographystyle{apsrev}
\bibliography{fermiball}

\begin{thebibliography}{74}
\expandafter\ifx\csname natexlab\endcsname\relax\def\natexlab#1{#1}\fi
\expandafter\ifx\csname bibnamefont\endcsname\relax
  \def\bibnamefont#1{#1}\fi
\expandafter\ifx\csname bibfnamefont\endcsname\relax
  \def\bibfnamefont#1{#1}\fi
\expandafter\ifx\csname citenamefont\endcsname\relax
  \def\citenamefont#1{#1}\fi
\expandafter\ifx\csname url\endcsname\relax
  \def\url#1{\texttt{#1}}\fi
\expandafter\ifx\csname urlprefix\endcsname\relax\def\urlprefix{URL }\fi
\providecommand{\bibinfo}[2]{#2}
\providecommand{\eprint}[2][]{\url{#2}}

\bibitem[{\citenamefont{Lu et~al.}(2021)\citenamefont{Lu, Lu, Ma, Li, Huang,
  Jia, Ma, Liu, Zhang, Yu et~al.}}]{Lu2021.04.15.439963}
\bibinfo{author}{\bibfnamefont{Y.}~\bibnamefont{Lu}},
  \bibinfo{author}{\bibfnamefont{Y.}~\bibnamefont{Lu}},
  \bibinfo{author}{\bibfnamefont{J.}~\bibnamefont{Ma}},
  \bibinfo{author}{\bibfnamefont{J.}~\bibnamefont{Li}},
  \bibinfo{author}{\bibfnamefont{X.}~\bibnamefont{Huang}},
  \bibinfo{author}{\bibfnamefont{Q.}~\bibnamefont{Jia}},
  \bibinfo{author}{\bibfnamefont{D.}~\bibnamefont{Ma}},
  \bibinfo{author}{\bibfnamefont{M.}~\bibnamefont{Liu}},
  \bibinfo{author}{\bibfnamefont{H.}~\bibnamefont{Zhang}},
  \bibinfo{author}{\bibfnamefont{X.}~\bibnamefont{Yu}}, \bibnamefont{et~al.},
  \bibinfo{journal}{bioRxiv}  (\bibinfo{year}{2021}),
  \urlprefix\url{https://www.biorxiv.org/content/early/2021/04/20/2021.04.15.439963}.

\bibitem[{\citenamefont{Narayanan et~al.}(2017)\citenamefont{Narayanan, Meriin,
  Sherman, and Ciss{\'e}}}]{Narayanan148395}
\bibinfo{author}{\bibfnamefont{A.}~\bibnamefont{Narayanan}},
  \bibinfo{author}{\bibfnamefont{A.~B.} \bibnamefont{Meriin}},
  \bibinfo{author}{\bibfnamefont{M.~Y.} \bibnamefont{Sherman}},
  \bibnamefont{and} \bibinfo{author}{\bibfnamefont{I.~I.}
  \bibnamefont{Ciss{\'e}}}, \bibinfo{journal}{bioRxiv}  (\bibinfo{year}{2017}),
  \urlprefix\url{https://www.biorxiv.org/content/early/2017/06/19/148395}.

\bibitem[{\citenamefont{{Blume}}(1966)}]{1966PhRv..141..517B}
\bibinfo{author}{\bibfnamefont{M.}~\bibnamefont{{Blume}}},
  \bibinfo{journal}{Physical Review} \textbf{\bibinfo{volume}{141}},
  \bibinfo{pages}{517} (\bibinfo{year}{1966}).

\bibitem[{\citenamefont{{Sachdev}}(2011)}]{2011qpt..book.....S}
\bibinfo{author}{\bibfnamefont{S.}~\bibnamefont{{Sachdev}}},
  \emph{\bibinfo{title}{{Quantum Phase Transitions}}} (\bibinfo{year}{2011}).

\bibitem[{\citenamefont{Sato}(1981)}]{Sato:1980yn}
\bibinfo{author}{\bibfnamefont{K.}~\bibnamefont{Sato}}, \bibinfo{journal}{Mon.
  Not. Roy. Astron. Soc.} \textbf{\bibinfo{volume}{195}}, \bibinfo{pages}{467}
  (\bibinfo{year}{1981}).

\bibitem[{\citenamefont{Kuzmin et~al.}(1985)\citenamefont{Kuzmin, Rubakov, and
  Shaposhnikov}}]{Kuzmin:1985mm}
\bibinfo{author}{\bibfnamefont{V.~A.} \bibnamefont{Kuzmin}},
  \bibinfo{author}{\bibfnamefont{V.~A.} \bibnamefont{Rubakov}},
  \bibnamefont{and} \bibinfo{author}{\bibfnamefont{M.~E.}
  \bibnamefont{Shaposhnikov}}, \bibinfo{journal}{Phys. Lett. B}
  \textbf{\bibinfo{volume}{155}}, \bibinfo{pages}{36} (\bibinfo{year}{1985}).

\bibitem[{\citenamefont{Joyce et~al.}(1996)\citenamefont{Joyce, Prokopec, and
  Turok}}]{Joyce:1994zt}
\bibinfo{author}{\bibfnamefont{M.}~\bibnamefont{Joyce}},
  \bibinfo{author}{\bibfnamefont{T.}~\bibnamefont{Prokopec}}, \bibnamefont{and}
  \bibinfo{author}{\bibfnamefont{N.}~\bibnamefont{Turok}},
  \bibinfo{journal}{Phys. Rev. D} \textbf{\bibinfo{volume}{53}},
  \bibinfo{pages}{2958} (\bibinfo{year}{1996}), \eprint{hep-ph/9410282}.

\bibitem[{\citenamefont{Joyce et~al.}(1995)\citenamefont{Joyce, Prokopec, and
  Turok}}]{Joyce:1994fu}
\bibinfo{author}{\bibfnamefont{M.}~\bibnamefont{Joyce}},
  \bibinfo{author}{\bibfnamefont{T.}~\bibnamefont{Prokopec}}, \bibnamefont{and}
  \bibinfo{author}{\bibfnamefont{N.}~\bibnamefont{Turok}},
  \bibinfo{journal}{Phys. Rev. Lett.} \textbf{\bibinfo{volume}{75}},
  \bibinfo{pages}{1695} (\bibinfo{year}{1995}), \bibinfo{note}{[Erratum:
  Phys.Rev.Lett. 75, 3375 (1995)]}, \eprint{hep-ph/9408339}.

\bibitem[{\citenamefont{Cohen et~al.}(1993)\citenamefont{Cohen, Kaplan, and
  Nelson}}]{Cohen:1993nk}
\bibinfo{author}{\bibfnamefont{A.~G.} \bibnamefont{Cohen}},
  \bibinfo{author}{\bibfnamefont{D.~B.} \bibnamefont{Kaplan}},
  \bibnamefont{and} \bibinfo{author}{\bibfnamefont{A.~E.}
  \bibnamefont{Nelson}}, \bibinfo{journal}{Ann. Rev. Nucl. Part. Sci.}
  \textbf{\bibinfo{volume}{43}}, \bibinfo{pages}{27} (\bibinfo{year}{1993}),
  \eprint{hep-ph/9302210}.

\bibitem[{\citenamefont{Morrissey and Ramsey-Musolf}(2012)}]{Morrissey:2012db}
\bibinfo{author}{\bibfnamefont{D.~E.} \bibnamefont{Morrissey}}
  \bibnamefont{and} \bibinfo{author}{\bibfnamefont{M.~J.}
  \bibnamefont{Ramsey-Musolf}}, \bibinfo{journal}{New J. Phys.}
  \textbf{\bibinfo{volume}{14}}, \bibinfo{pages}{125003}
  (\bibinfo{year}{2012}), \eprint{1206.2942}.

\bibitem[{\citenamefont{Baker and Kopp}(2017)}]{Baker:2016xzo}
\bibinfo{author}{\bibfnamefont{M.~J.} \bibnamefont{Baker}} \bibnamefont{and}
  \bibinfo{author}{\bibfnamefont{J.}~\bibnamefont{Kopp}},
  \bibinfo{journal}{Phys. Rev. Lett.} \textbf{\bibinfo{volume}{119}},
  \bibinfo{pages}{061801} (\bibinfo{year}{2017}), \eprint{1608.07578}.

\bibitem[{\citenamefont{Baker et~al.}(2020)\citenamefont{Baker, Kopp, and
  Long}}]{Baker:2019ndr}
\bibinfo{author}{\bibfnamefont{M.~J.} \bibnamefont{Baker}},
  \bibinfo{author}{\bibfnamefont{J.}~\bibnamefont{Kopp}}, \bibnamefont{and}
  \bibinfo{author}{\bibfnamefont{A.~J.} \bibnamefont{Long}},
  \bibinfo{journal}{Phys. Rev. Lett.} \textbf{\bibinfo{volume}{125}},
  \bibinfo{pages}{151102} (\bibinfo{year}{2020}), \eprint{1912.02830}.

\bibitem[{\citenamefont{Chway et~al.}(2020)\citenamefont{Chway, Jung, and
  Shin}}]{Chway:2019kft}
\bibinfo{author}{\bibfnamefont{D.}~\bibnamefont{Chway}},
  \bibinfo{author}{\bibfnamefont{T.~H.} \bibnamefont{Jung}}, \bibnamefont{and}
  \bibinfo{author}{\bibfnamefont{C.~S.} \bibnamefont{Shin}},
  \bibinfo{journal}{Phys. Rev. D} \textbf{\bibinfo{volume}{101}},
  \bibinfo{pages}{095019} (\bibinfo{year}{2020}), \eprint{1912.04238}.

\bibitem[{\citenamefont{Azatov et~al.}(2021)\citenamefont{Azatov,
  Vanvlasselaer, and Yin}}]{Azatov:2021ifm}
\bibinfo{author}{\bibfnamefont{A.}~\bibnamefont{Azatov}},
  \bibinfo{author}{\bibfnamefont{M.}~\bibnamefont{Vanvlasselaer}},
  \bibnamefont{and} \bibinfo{author}{\bibfnamefont{W.}~\bibnamefont{Yin}},
  \bibinfo{journal}{JHEP} \textbf{\bibinfo{volume}{03}}, \bibinfo{pages}{288}
  (\bibinfo{year}{2021}), \eprint{2101.05721}.

\bibitem[{\citenamefont{Witten}(1984)}]{Witten:1984rs}
\bibinfo{author}{\bibfnamefont{E.}~\bibnamefont{Witten}},
  \bibinfo{journal}{Phys. Rev. D} \textbf{\bibinfo{volume}{30}},
  \bibinfo{pages}{272} (\bibinfo{year}{1984}).

\bibitem[{\citenamefont{Krylov et~al.}(2013)\citenamefont{Krylov, Levin, and
  Rubakov}}]{Krylov:2013qe}
\bibinfo{author}{\bibfnamefont{E.}~\bibnamefont{Krylov}},
  \bibinfo{author}{\bibfnamefont{A.}~\bibnamefont{Levin}}, \bibnamefont{and}
  \bibinfo{author}{\bibfnamefont{V.}~\bibnamefont{Rubakov}},
  \bibinfo{journal}{Phys. Rev. D} \textbf{\bibinfo{volume}{87}},
  \bibinfo{pages}{083528} (\bibinfo{year}{2013}), \eprint{1301.0354}.

\bibitem[{\citenamefont{Huang and Li}(2017)}]{Huang:2017kzu}
\bibinfo{author}{\bibfnamefont{F.~P.} \bibnamefont{Huang}} \bibnamefont{and}
  \bibinfo{author}{\bibfnamefont{C.~S.} \bibnamefont{Li}},
  \bibinfo{journal}{Phys. Rev. D} \textbf{\bibinfo{volume}{96}},
  \bibinfo{pages}{095028} (\bibinfo{year}{2017}), \eprint{1709.09691}.

\bibitem[{\citenamefont{Bai and Long}(2018)}]{Bai:2018vik}
\bibinfo{author}{\bibfnamefont{Y.}~\bibnamefont{Bai}} \bibnamefont{and}
  \bibinfo{author}{\bibfnamefont{A.~J.} \bibnamefont{Long}},
  \bibinfo{journal}{JHEP} \textbf{\bibinfo{volume}{06}}, \bibinfo{pages}{072}
  (\bibinfo{year}{2018}), \eprint{1804.10249}.

\bibitem[{\citenamefont{Bai et~al.}(2019)\citenamefont{Bai, Long, and
  Lu}}]{Bai:2018dxf}
\bibinfo{author}{\bibfnamefont{Y.}~\bibnamefont{Bai}},
  \bibinfo{author}{\bibfnamefont{A.~J.} \bibnamefont{Long}}, \bibnamefont{and}
  \bibinfo{author}{\bibfnamefont{S.}~\bibnamefont{Lu}}, \bibinfo{journal}{Phys.
  Rev. D} \textbf{\bibinfo{volume}{99}}, \bibinfo{pages}{055047}
  (\bibinfo{year}{2019}), \eprint{1810.04360}.

\bibitem[{\citenamefont{Hong et~al.}(2020)\citenamefont{Hong, Jung, and
  Xie}}]{Hong:2020est}
\bibinfo{author}{\bibfnamefont{J.-P.} \bibnamefont{Hong}},
  \bibinfo{author}{\bibfnamefont{S.}~\bibnamefont{Jung}}, \bibnamefont{and}
  \bibinfo{author}{\bibfnamefont{K.-P.} \bibnamefont{Xie}},
  \bibinfo{journal}{Phys. Rev. D} \textbf{\bibinfo{volume}{102}},
  \bibinfo{pages}{075028} (\bibinfo{year}{2020}), \eprint{2008.04430}.

\bibitem[{\citenamefont{Asadi et~al.}(2021)\citenamefont{Asadi, Kramer, Kuflik,
  Ridgway, Slatyer, and Smirnov}}]{Asadi:2021yml}
\bibinfo{author}{\bibfnamefont{P.}~\bibnamefont{Asadi}},
  \bibinfo{author}{\bibfnamefont{E.~D.} \bibnamefont{Kramer}},
  \bibinfo{author}{\bibfnamefont{E.}~\bibnamefont{Kuflik}},
  \bibinfo{author}{\bibfnamefont{G.~W.} \bibnamefont{Ridgway}},
  \bibinfo{author}{\bibfnamefont{T.~R.} \bibnamefont{Slatyer}},
  \bibnamefont{and} \bibinfo{author}{\bibfnamefont{J.}~\bibnamefont{Smirnov}},
  \bibinfo{journal}{Phys. Rev. Lett.} \textbf{\bibinfo{volume}{127}},
  \bibinfo{pages}{211101} (\bibinfo{year}{2021}), \eprint{2103.09822}.

\bibitem[{\citenamefont{Marfatia and
  Tseng}(2021{\natexlab{a}})}]{Marfatia:2021twj}
\bibinfo{author}{\bibfnamefont{D.}~\bibnamefont{Marfatia}} \bibnamefont{and}
  \bibinfo{author}{\bibfnamefont{P.-Y.} \bibnamefont{Tseng}},
  \bibinfo{journal}{JHEP} \textbf{\bibinfo{volume}{11}}, \bibinfo{pages}{068}
  (\bibinfo{year}{2021}{\natexlab{a}}), \eprint{2107.00859}.

\bibitem[{\citenamefont{Crawford and Schramm}(1982)}]{Crawford:1982yz}
\bibinfo{author}{\bibfnamefont{M.}~\bibnamefont{Crawford}} \bibnamefont{and}
  \bibinfo{author}{\bibfnamefont{D.~N.} \bibnamefont{Schramm}},
  \bibinfo{journal}{Nature} \textbf{\bibinfo{volume}{298}},
  \bibinfo{pages}{538} (\bibinfo{year}{1982}).

\bibitem[{\citenamefont{Hawking et~al.}(1982)\citenamefont{Hawking, Moss, and
  Stewart}}]{Hawking:1982ga}
\bibinfo{author}{\bibfnamefont{S.~W.} \bibnamefont{Hawking}},
  \bibinfo{author}{\bibfnamefont{I.~G.} \bibnamefont{Moss}}, \bibnamefont{and}
  \bibinfo{author}{\bibfnamefont{J.~M.} \bibnamefont{Stewart}},
  \bibinfo{journal}{Phys. Rev. D} \textbf{\bibinfo{volume}{26}},
  \bibinfo{pages}{2681} (\bibinfo{year}{1982}).

\bibitem[{\citenamefont{La and Steinhardt}(1989)}]{La:1989st}
\bibinfo{author}{\bibfnamefont{D.}~\bibnamefont{La}} \bibnamefont{and}
  \bibinfo{author}{\bibfnamefont{P.~J.} \bibnamefont{Steinhardt}},
  \bibinfo{journal}{Phys. Lett. B} \textbf{\bibinfo{volume}{220}},
  \bibinfo{pages}{375} (\bibinfo{year}{1989}).

\bibitem[{\citenamefont{Moss}(1994)}]{Moss:1994iq}
\bibinfo{author}{\bibfnamefont{I.~G.} \bibnamefont{Moss}},
  \bibinfo{journal}{Phys. Rev. D} \textbf{\bibinfo{volume}{50}},
  \bibinfo{pages}{676} (\bibinfo{year}{1994}).

\bibitem[{\citenamefont{Konoplich et~al.}(1998)\citenamefont{Konoplich, Rubin,
  Sakharov, and Khlopov}}]{konoplich1998formation}
\bibinfo{author}{\bibfnamefont{R.}~\bibnamefont{Konoplich}},
  \bibinfo{author}{\bibfnamefont{S.}~\bibnamefont{Rubin}},
  \bibinfo{author}{\bibfnamefont{A.}~\bibnamefont{Sakharov}}, \bibnamefont{and}
  \bibinfo{author}{\bibfnamefont{M.~Y.} \bibnamefont{Khlopov}},
  \bibinfo{journal}{Astronomy Letters} \textbf{\bibinfo{volume}{24}},
  \bibinfo{pages}{413} (\bibinfo{year}{1998}).

\bibitem[{\citenamefont{Konoplich et~al.}(1999)\citenamefont{Konoplich, Rubin,
  Sakharov, and Khlopov}}]{Konoplich:1999qq}
\bibinfo{author}{\bibfnamefont{R.~V.} \bibnamefont{Konoplich}},
  \bibinfo{author}{\bibfnamefont{S.~G.} \bibnamefont{Rubin}},
  \bibinfo{author}{\bibfnamefont{A.~S.} \bibnamefont{Sakharov}},
  \bibnamefont{and} \bibinfo{author}{\bibfnamefont{M.~Y.}
  \bibnamefont{Khlopov}}, \bibinfo{journal}{Phys. Atom. Nucl.}
  \textbf{\bibinfo{volume}{62}}, \bibinfo{pages}{1593} (\bibinfo{year}{1999}).

\bibitem[{\citenamefont{Kodama et~al.}(1982)\citenamefont{Kodama, Sasaki, and
  Sato}}]{Kodama:1982sf}
\bibinfo{author}{\bibfnamefont{H.}~\bibnamefont{Kodama}},
  \bibinfo{author}{\bibfnamefont{M.}~\bibnamefont{Sasaki}}, \bibnamefont{and}
  \bibinfo{author}{\bibfnamefont{K.}~\bibnamefont{Sato}},
  \bibinfo{journal}{Prog. Theor. Phys.} \textbf{\bibinfo{volume}{68}},
  \bibinfo{pages}{1979} (\bibinfo{year}{1982}).

\bibitem[{\citenamefont{Lewicki and Vaskonen}(2020)}]{Lewicki:2019gmv}
\bibinfo{author}{\bibfnamefont{M.}~\bibnamefont{Lewicki}} \bibnamefont{and}
  \bibinfo{author}{\bibfnamefont{V.}~\bibnamefont{Vaskonen}},
  \bibinfo{journal}{Phys. Dark Univ.} \textbf{\bibinfo{volume}{30}},
  \bibinfo{pages}{100672} (\bibinfo{year}{2020}), \eprint{1912.00997}.

\bibitem[{\citenamefont{Kusenko et~al.}(2020)\citenamefont{Kusenko, Sasaki,
  Sugiyama, Takada, Takhistov, and Vitagliano}}]{Kusenko:2020pcg}
\bibinfo{author}{\bibfnamefont{A.}~\bibnamefont{Kusenko}},
  \bibinfo{author}{\bibfnamefont{M.}~\bibnamefont{Sasaki}},
  \bibinfo{author}{\bibfnamefont{S.}~\bibnamefont{Sugiyama}},
  \bibinfo{author}{\bibfnamefont{M.}~\bibnamefont{Takada}},
  \bibinfo{author}{\bibfnamefont{V.}~\bibnamefont{Takhistov}},
  \bibnamefont{and}
  \bibinfo{author}{\bibfnamefont{E.}~\bibnamefont{Vitagliano}},
  \bibinfo{journal}{Phys. Rev. Lett.} \textbf{\bibinfo{volume}{125}},
  \bibinfo{pages}{181304} (\bibinfo{year}{2020}), \eprint{2001.09160}.

\bibitem[{\citenamefont{Gross et~al.}(2021)\citenamefont{Gross, Landini,
  Strumia, and Teresi}}]{Gross:2021qgx}
\bibinfo{author}{\bibfnamefont{C.}~\bibnamefont{Gross}},
  \bibinfo{author}{\bibfnamefont{G.}~\bibnamefont{Landini}},
  \bibinfo{author}{\bibfnamefont{A.}~\bibnamefont{Strumia}}, \bibnamefont{and}
  \bibinfo{author}{\bibfnamefont{D.}~\bibnamefont{Teresi}},
  \bibinfo{journal}{JHEP} \textbf{\bibinfo{volume}{09}}, \bibinfo{pages}{033}
  (\bibinfo{year}{2021}), \eprint{2105.02840}.

\bibitem[{\citenamefont{Baker et~al.}(2021{\natexlab{a}})\citenamefont{Baker,
  Breitbach, Kopp, and Mittnacht}}]{Baker:2021nyl}
\bibinfo{author}{\bibfnamefont{M.~J.} \bibnamefont{Baker}},
  \bibinfo{author}{\bibfnamefont{M.}~\bibnamefont{Breitbach}},
  \bibinfo{author}{\bibfnamefont{J.}~\bibnamefont{Kopp}}, \bibnamefont{and}
  \bibinfo{author}{\bibfnamefont{L.}~\bibnamefont{Mittnacht}}
  (\bibinfo{year}{2021}{\natexlab{a}}), \eprint{2105.07481}.

\bibitem[{\citenamefont{Baker et~al.}(2021{\natexlab{b}})\citenamefont{Baker,
  Breitbach, Kopp, and Mittnacht}}]{Baker:2021sno}
\bibinfo{author}{\bibfnamefont{M.~J.} \bibnamefont{Baker}},
  \bibinfo{author}{\bibfnamefont{M.}~\bibnamefont{Breitbach}},
  \bibinfo{author}{\bibfnamefont{J.}~\bibnamefont{Kopp}}, \bibnamefont{and}
  \bibinfo{author}{\bibfnamefont{L.}~\bibnamefont{Mittnacht}}
  (\bibinfo{year}{2021}{\natexlab{b}}), \eprint{2110.00005}.

\bibitem[{\citenamefont{Kawana and Xie}(2022)}]{Kawana:2021tde}
\bibinfo{author}{\bibfnamefont{K.}~\bibnamefont{Kawana}} \bibnamefont{and}
  \bibinfo{author}{\bibfnamefont{K.-P.} \bibnamefont{Xie}},
  \bibinfo{journal}{Phys. Lett. B} \textbf{\bibinfo{volume}{824}},
  \bibinfo{pages}{136791} (\bibinfo{year}{2022}), \eprint{2106.00111}.

\bibitem[{\citenamefont{Marfatia and
  Tseng}(2021{\natexlab{b}})}]{Marfatia:2021hcp}
\bibinfo{author}{\bibfnamefont{D.}~\bibnamefont{Marfatia}} \bibnamefont{and}
  \bibinfo{author}{\bibfnamefont{P.-Y.} \bibnamefont{Tseng}}
  (\bibinfo{year}{2021}{\natexlab{b}}), \eprint{2112.14588}.

\bibitem[{\citenamefont{Huang and Xie}(2022)}]{Huang:2022him}
\bibinfo{author}{\bibfnamefont{P.}~\bibnamefont{Huang}} \bibnamefont{and}
  \bibinfo{author}{\bibfnamefont{K.-P.} \bibnamefont{Xie}}
  (\bibinfo{year}{2022}), \eprint{2201.07243}.

\bibitem[{\citenamefont{Liu et~al.}(2021)\citenamefont{Liu, Bian, Cai, Guo, and
  Wang}}]{Liu:2021svg}
\bibinfo{author}{\bibfnamefont{J.}~\bibnamefont{Liu}},
  \bibinfo{author}{\bibfnamefont{L.}~\bibnamefont{Bian}},
  \bibinfo{author}{\bibfnamefont{R.-G.} \bibnamefont{Cai}},
  \bibinfo{author}{\bibfnamefont{Z.-K.} \bibnamefont{Guo}}, \bibnamefont{and}
  \bibinfo{author}{\bibfnamefont{S.-J.} \bibnamefont{Wang}}
  (\bibinfo{year}{2021}), \eprint{2106.05637}.

\bibitem[{\citenamefont{Davoudiasl et~al.}(2021)\citenamefont{Davoudiasl,
  Denton, and Gehrlein}}]{Davoudiasl:2021olb}
\bibinfo{author}{\bibfnamefont{H.}~\bibnamefont{Davoudiasl}},
  \bibinfo{author}{\bibfnamefont{P.~B.} \bibnamefont{Denton}},
  \bibnamefont{and} \bibinfo{author}{\bibfnamefont{J.}~\bibnamefont{Gehrlein}}
  (\bibinfo{year}{2021}), \eprint{2109.01678}.

\bibitem[{\citenamefont{Jung and Okui}(2021)}]{Jung:2021mku}
\bibinfo{author}{\bibfnamefont{T.~H.} \bibnamefont{Jung}} \bibnamefont{and}
  \bibinfo{author}{\bibfnamefont{T.}~\bibnamefont{Okui}}
  (\bibinfo{year}{2021}), \eprint{2110.04271}.

\bibitem[{\citenamefont{Hashino et~al.}(2021)\citenamefont{Hashino, Kanemura,
  and Takahashi}}]{Hashino:2021qoq}
\bibinfo{author}{\bibfnamefont{K.}~\bibnamefont{Hashino}},
  \bibinfo{author}{\bibfnamefont{S.}~\bibnamefont{Kanemura}}, \bibnamefont{and}
  \bibinfo{author}{\bibfnamefont{T.}~\bibnamefont{Takahashi}}
  (\bibinfo{year}{2021}), \eprint{2111.13099}.

\bibitem[{\citenamefont{Maeso et~al.}(2021)\citenamefont{Maeso, Marzola,
  Raidal, Vaskonen, and Veerm\"ae}}]{Maeso:2021xvl}
\bibinfo{author}{\bibfnamefont{D.~N.} \bibnamefont{Maeso}},
  \bibinfo{author}{\bibfnamefont{L.}~\bibnamefont{Marzola}},
  \bibinfo{author}{\bibfnamefont{M.}~\bibnamefont{Raidal}},
  \bibinfo{author}{\bibfnamefont{V.}~\bibnamefont{Vaskonen}}, \bibnamefont{and}
  \bibinfo{author}{\bibfnamefont{H.}~\bibnamefont{Veerm\"ae}}
  (\bibinfo{year}{2021}), \eprint{2112.01505}.

\bibitem[{\citenamefont{Xue et~al.}(2021)}]{Xue:2021gyq}
\bibinfo{author}{\bibfnamefont{X.}~\bibnamefont{Xue}} \bibnamefont{et~al.},
  \bibinfo{journal}{Phys. Rev. Lett.} \textbf{\bibinfo{volume}{127}},
  \bibinfo{pages}{251303} (\bibinfo{year}{2021}), \eprint{2110.03096}.

\bibitem[{\citenamefont{Arzoumanian et~al.}(2020)}]{NANOGrav:2020bcs}
\bibinfo{author}{\bibfnamefont{Z.}~\bibnamefont{Arzoumanian}}
  \bibnamefont{et~al.} (\bibinfo{collaboration}{NANOGrav}),
  \bibinfo{journal}{Astrophys. J. Lett.} \textbf{\bibinfo{volume}{905}},
  \bibinfo{pages}{L34} (\bibinfo{year}{2020}), \eprint{2009.04496}.

\bibitem[{\citenamefont{Romero et~al.}(2021)\citenamefont{Romero, Martinovic,
  Callister, Guo, Mart\'\i{}nez, Sakellariadou, Yang, and
  Zhao}}]{Romero:2021kby}
\bibinfo{author}{\bibfnamefont{A.}~\bibnamefont{Romero}},
  \bibinfo{author}{\bibfnamefont{K.}~\bibnamefont{Martinovic}},
  \bibinfo{author}{\bibfnamefont{T.~A.} \bibnamefont{Callister}},
  \bibinfo{author}{\bibfnamefont{H.-K.} \bibnamefont{Guo}},
  \bibinfo{author}{\bibfnamefont{M.}~\bibnamefont{Mart\'\i{}nez}},
  \bibinfo{author}{\bibfnamefont{M.}~\bibnamefont{Sakellariadou}},
  \bibinfo{author}{\bibfnamefont{F.-W.} \bibnamefont{Yang}}, \bibnamefont{and}
  \bibinfo{author}{\bibfnamefont{Y.}~\bibnamefont{Zhao}},
  \bibinfo{journal}{Phys. Rev. Lett.} \textbf{\bibinfo{volume}{126}},
  \bibinfo{pages}{151301} (\bibinfo{year}{2021}), \eprint{2102.01714}.

\bibitem[{\citenamefont{Caprini et~al.}(2016)}]{Caprini:2015zlo}
\bibinfo{author}{\bibfnamefont{C.}~\bibnamefont{Caprini}} \bibnamefont{et~al.},
  \bibinfo{journal}{JCAP} \textbf{\bibinfo{volume}{04}}, \bibinfo{pages}{001}
  (\bibinfo{year}{2016}), \eprint{1512.06239}.

\bibitem[{\citenamefont{Caprini et~al.}(2020)}]{Caprini:2019egz}
\bibinfo{author}{\bibfnamefont{C.}~\bibnamefont{Caprini}} \bibnamefont{et~al.},
  \bibinfo{journal}{JCAP} \textbf{\bibinfo{volume}{03}}, \bibinfo{pages}{024}
  (\bibinfo{year}{2020}), \eprint{1910.13125}.

\bibitem[{\citenamefont{Liang et~al.}(2022)\citenamefont{Liang, Hu, Jiang,
  Cheng, Zhang, and Mei}}]{Liang:2021bde}
\bibinfo{author}{\bibfnamefont{Z.-C.} \bibnamefont{Liang}},
  \bibinfo{author}{\bibfnamefont{Y.-M.} \bibnamefont{Hu}},
  \bibinfo{author}{\bibfnamefont{Y.}~\bibnamefont{Jiang}},
  \bibinfo{author}{\bibfnamefont{J.}~\bibnamefont{Cheng}},
  \bibinfo{author}{\bibfnamefont{J.-d.} \bibnamefont{Zhang}}, \bibnamefont{and}
  \bibinfo{author}{\bibfnamefont{J.}~\bibnamefont{Mei}},
  \bibinfo{journal}{Phys. Rev. D} \textbf{\bibinfo{volume}{105}},
  \bibinfo{pages}{022001} (\bibinfo{year}{2022}), \eprint{2107.08643}.

\bibitem[{\citenamefont{Hogan}(1986)}]{Hogan:1986qda}
\bibinfo{author}{\bibfnamefont{C.~J.} \bibnamefont{Hogan}},
  \bibinfo{journal}{Mon. Not. Roy. Astron. Soc.}
  \textbf{\bibinfo{volume}{218}}, \bibinfo{pages}{629} (\bibinfo{year}{1986}).

\bibitem[{\citenamefont{Maggiore}(2000)}]{Maggiore:1999vm}
\bibinfo{author}{\bibfnamefont{M.}~\bibnamefont{Maggiore}},
  \bibinfo{journal}{Phys. Rept.} \textbf{\bibinfo{volume}{331}},
  \bibinfo{pages}{283} (\bibinfo{year}{2000}), \eprint{gr-qc/9909001}.

\bibitem[{\citenamefont{Kawana}(2022)}]{Kawana:2022fum}
\bibinfo{author}{\bibfnamefont{K.}~\bibnamefont{Kawana}}
  (\bibinfo{year}{2022}), \eprint{2201.00560}.

\bibitem[{\citenamefont{Callan and Coleman}(1977)}]{Callan:1977pt}
\bibinfo{author}{\bibfnamefont{C.~G.} \bibnamefont{Callan}, \bibfnamefont{Jr.}}
  \bibnamefont{and} \bibinfo{author}{\bibfnamefont{S.~R.}
  \bibnamefont{Coleman}}, \bibinfo{journal}{Phys. Rev. D}
  \textbf{\bibinfo{volume}{16}}, \bibinfo{pages}{1762} (\bibinfo{year}{1977}).

\bibitem[{\citenamefont{Guth and Tye}(1980)}]{Guth:1979bh}
\bibinfo{author}{\bibfnamefont{A.~H.} \bibnamefont{Guth}} \bibnamefont{and}
  \bibinfo{author}{\bibfnamefont{S.~H.~H.} \bibnamefont{Tye}},
  \bibinfo{journal}{Phys. Rev. Lett.} \textbf{\bibinfo{volume}{44}},
  \bibinfo{pages}{631} (\bibinfo{year}{1980}), \bibinfo{note}{[Erratum:
  Phys.Rev.Lett. 44, 963 (1980)]}.

\bibitem[{\citenamefont{Guth and Weinberg}(1981)}]{Guth:1981uk}
\bibinfo{author}{\bibfnamefont{A.~H.} \bibnamefont{Guth}} \bibnamefont{and}
  \bibinfo{author}{\bibfnamefont{E.~J.} \bibnamefont{Weinberg}},
  \bibinfo{journal}{Phys. Rev. D} \textbf{\bibinfo{volume}{23}},
  \bibinfo{pages}{876} (\bibinfo{year}{1981}).

\bibitem[{\citenamefont{Affleck}(1981)}]{Affleck:1980ac}
\bibinfo{author}{\bibfnamefont{I.}~\bibnamefont{Affleck}},
  \bibinfo{journal}{Phys. Rev. Lett.} \textbf{\bibinfo{volume}{46}},
  \bibinfo{pages}{388} (\bibinfo{year}{1981}).

\bibitem[{\citenamefont{Dine et~al.}(1992)\citenamefont{Dine, Leigh, Huet,
  Linde, and Linde}}]{Dine:1992wr}
\bibinfo{author}{\bibfnamefont{M.}~\bibnamefont{Dine}},
  \bibinfo{author}{\bibfnamefont{R.~G.} \bibnamefont{Leigh}},
  \bibinfo{author}{\bibfnamefont{P.~Y.} \bibnamefont{Huet}},
  \bibinfo{author}{\bibfnamefont{A.~D.} \bibnamefont{Linde}}, \bibnamefont{and}
  \bibinfo{author}{\bibfnamefont{D.~A.} \bibnamefont{Linde}},
  \bibinfo{journal}{Phys. Rev. D} \textbf{\bibinfo{volume}{46}},
  \bibinfo{pages}{550} (\bibinfo{year}{1992}), \eprint{hep-ph/9203203}.

\bibitem[{\citenamefont{Espinosa et~al.}(2010)\citenamefont{Espinosa,
  Konstandin, No, and Servant}}]{Espinosa:2010hh}
\bibinfo{author}{\bibfnamefont{J.~R.} \bibnamefont{Espinosa}},
  \bibinfo{author}{\bibfnamefont{T.}~\bibnamefont{Konstandin}},
  \bibinfo{author}{\bibfnamefont{J.~M.} \bibnamefont{No}}, \bibnamefont{and}
  \bibinfo{author}{\bibfnamefont{G.}~\bibnamefont{Servant}},
  \bibinfo{journal}{JCAP} \textbf{\bibinfo{volume}{06}}, \bibinfo{pages}{028}
  (\bibinfo{year}{2010}), \eprint{1004.4187}.

\bibitem[{\citenamefont{Megevand and Ramirez}(2017)}]{Megevand:2016lpr}
\bibinfo{author}{\bibfnamefont{A.}~\bibnamefont{Megevand}} \bibnamefont{and}
  \bibinfo{author}{\bibfnamefont{S.}~\bibnamefont{Ramirez}},
  \bibinfo{journal}{Nucl. Phys. B} \textbf{\bibinfo{volume}{919}},
  \bibinfo{pages}{74} (\bibinfo{year}{2017}), \eprint{1611.05853}.

\bibitem[{\citenamefont{Ellis et~al.}(2019)\citenamefont{Ellis, Lewicki, and
  No}}]{Ellis:2018mja}
\bibinfo{author}{\bibfnamefont{J.}~\bibnamefont{Ellis}},
  \bibinfo{author}{\bibfnamefont{M.}~\bibnamefont{Lewicki}}, \bibnamefont{and}
  \bibinfo{author}{\bibfnamefont{J.~M.} \bibnamefont{No}},
  \bibinfo{journal}{JCAP} \textbf{\bibinfo{volume}{04}}, \bibinfo{pages}{003}
  (\bibinfo{year}{2019}), \eprint{1809.08242}.

\bibitem[{\citenamefont{Wang et~al.}(2020)\citenamefont{Wang, Huang, and
  Zhang}}]{Wang:2020jrd}
\bibinfo{author}{\bibfnamefont{X.}~\bibnamefont{Wang}},
  \bibinfo{author}{\bibfnamefont{F.~P.} \bibnamefont{Huang}}, \bibnamefont{and}
  \bibinfo{author}{\bibfnamefont{X.}~\bibnamefont{Zhang}},
  \bibinfo{journal}{JCAP} \textbf{\bibinfo{volume}{05}}, \bibinfo{pages}{045}
  (\bibinfo{year}{2020}), \eprint{2003.08892}.

\bibitem[{\citenamefont{Bodeker and Moore}(2009)}]{Bodeker:2009qy}
\bibinfo{author}{\bibfnamefont{D.}~\bibnamefont{Bodeker}} \bibnamefont{and}
  \bibinfo{author}{\bibfnamefont{G.~D.} \bibnamefont{Moore}},
  \bibinfo{journal}{JCAP} \textbf{\bibinfo{volume}{05}}, \bibinfo{pages}{009}
  (\bibinfo{year}{2009}), \eprint{0903.4099}.

\bibitem[{\citenamefont{Bodeker and Moore}(2017)}]{Bodeker:2017cim}
\bibinfo{author}{\bibfnamefont{D.}~\bibnamefont{Bodeker}} \bibnamefont{and}
  \bibinfo{author}{\bibfnamefont{G.~D.} \bibnamefont{Moore}},
  \bibinfo{journal}{JCAP} \textbf{\bibinfo{volume}{05}}, \bibinfo{pages}{025}
  (\bibinfo{year}{2017}), \eprint{1703.08215}.

\bibitem[{\citenamefont{H\"oche et~al.}(2021)\citenamefont{H\"oche, Kozaczuk,
  Long, Turner, and Wang}}]{Hoche:2020ysm}
\bibinfo{author}{\bibfnamefont{S.}~\bibnamefont{H\"oche}},
  \bibinfo{author}{\bibfnamefont{J.}~\bibnamefont{Kozaczuk}},
  \bibinfo{author}{\bibfnamefont{A.~J.} \bibnamefont{Long}},
  \bibinfo{author}{\bibfnamefont{J.}~\bibnamefont{Turner}}, \bibnamefont{and}
  \bibinfo{author}{\bibfnamefont{Y.}~\bibnamefont{Wang}},
  \bibinfo{journal}{JCAP} \textbf{\bibinfo{volume}{03}}, \bibinfo{pages}{009}
  (\bibinfo{year}{2021}), \eprint{2007.10343}.

\bibitem[{\citenamefont{Azatov and Vanvlasselaer}(2021)}]{Azatov:2020ufh}
\bibinfo{author}{\bibfnamefont{A.}~\bibnamefont{Azatov}} \bibnamefont{and}
  \bibinfo{author}{\bibfnamefont{M.}~\bibnamefont{Vanvlasselaer}},
  \bibinfo{journal}{JCAP} \textbf{\bibinfo{volume}{01}}, \bibinfo{pages}{058}
  (\bibinfo{year}{2021}), \eprint{2010.02590}.

\bibitem[{\citenamefont{Gouttenoire et~al.}(2021)\citenamefont{Gouttenoire,
  Jinno, and Sala}}]{Gouttenoire:2021kjv}
\bibinfo{author}{\bibfnamefont{Y.}~\bibnamefont{Gouttenoire}},
  \bibinfo{author}{\bibfnamefont{R.}~\bibnamefont{Jinno}}, \bibnamefont{and}
  \bibinfo{author}{\bibfnamefont{F.}~\bibnamefont{Sala}}
  (\bibinfo{year}{2021}), \eprint{2112.07686}.

\bibitem[{\citenamefont{De~Luca et~al.}(2021)\citenamefont{De~Luca,
  Franciolini, and Riotto}}]{DeLuca:2021mlh}
\bibinfo{author}{\bibfnamefont{V.}~\bibnamefont{De~Luca}},
  \bibinfo{author}{\bibfnamefont{G.}~\bibnamefont{Franciolini}},
  \bibnamefont{and} \bibinfo{author}{\bibfnamefont{A.}~\bibnamefont{Riotto}},
  \bibinfo{journal}{Phys. Rev. D} \textbf{\bibinfo{volume}{104}},
  \bibinfo{pages}{123539} (\bibinfo{year}{2021}), \eprint{2110.04229}.

\bibitem[{\citenamefont{Arakawa et~al.}(2021)\citenamefont{Arakawa, Rajaraman,
  and Tait}}]{Arakawa:2021wgz}
\bibinfo{author}{\bibfnamefont{J.}~\bibnamefont{Arakawa}},
  \bibinfo{author}{\bibfnamefont{A.}~\bibnamefont{Rajaraman}},
  \bibnamefont{and} \bibinfo{author}{\bibfnamefont{T.~M.~P.}
  \bibnamefont{Tait}} (\bibinfo{year}{2021}), \eprint{2109.13941}.

\bibitem[{\citenamefont{Linde}(1983)}]{Linde:1981zj}
\bibinfo{author}{\bibfnamefont{A.~D.} \bibnamefont{Linde}},
  \bibinfo{journal}{Nucl. Phys. B} \textbf{\bibinfo{volume}{216}},
  \bibinfo{pages}{421} (\bibinfo{year}{1983}), \bibinfo{note}{[Erratum:
  Nucl.Phys.B 223, 544 (1983)]}.

\bibitem[{\citenamefont{Coleman}(1977)}]{Coleman:1977py}
\bibinfo{author}{\bibfnamefont{S.~R.} \bibnamefont{Coleman}},
  \bibinfo{journal}{Phys. Rev. D} \textbf{\bibinfo{volume}{15}},
  \bibinfo{pages}{2929} (\bibinfo{year}{1977}), \bibinfo{note}{[Erratum:
  Phys.Rev.D 16, 1248 (1977)]}.

\bibitem[{\citenamefont{Rintoul and Torquato}(1997)}]{rintoul1997precise}
\bibinfo{author}{\bibfnamefont{M.~D.} \bibnamefont{Rintoul}} \bibnamefont{and}
  \bibinfo{author}{\bibfnamefont{S.}~\bibnamefont{Torquato}},
  \bibinfo{journal}{Journal of physics a: mathematical and general}
  \textbf{\bibinfo{volume}{30}}, \bibinfo{pages}{L585} (\bibinfo{year}{1997}).

\bibitem[{\citenamefont{Turner et~al.}(1992)\citenamefont{Turner, Weinberg, and
  Widrow}}]{Turner:1992tz}
\bibinfo{author}{\bibfnamefont{M.~S.} \bibnamefont{Turner}},
  \bibinfo{author}{\bibfnamefont{E.~J.} \bibnamefont{Weinberg}},
  \bibnamefont{and} \bibinfo{author}{\bibfnamefont{L.~M.}
  \bibnamefont{Widrow}}, \bibinfo{journal}{Phys. Rev. D}
  \textbf{\bibinfo{volume}{46}}, \bibinfo{pages}{2384} (\bibinfo{year}{1992}).

\bibitem[{\citenamefont{Carr et~al.}(2017)\citenamefont{Carr, Raidal, Tenkanen,
  Vaskonen, and Veerm\"ae}}]{Carr:2017jsz}
\bibinfo{author}{\bibfnamefont{B.}~\bibnamefont{Carr}},
  \bibinfo{author}{\bibfnamefont{M.}~\bibnamefont{Raidal}},
  \bibinfo{author}{\bibfnamefont{T.}~\bibnamefont{Tenkanen}},
  \bibinfo{author}{\bibfnamefont{V.}~\bibnamefont{Vaskonen}}, \bibnamefont{and}
  \bibinfo{author}{\bibfnamefont{H.}~\bibnamefont{Veerm\"ae}},
  \bibinfo{journal}{Phys. Rev. D} \textbf{\bibinfo{volume}{96}},
  \bibinfo{pages}{023514} (\bibinfo{year}{2017}), \eprint{1705.05567}.

\bibitem[{\citenamefont{Green and Kavanagh}(2021)}]{Green:2020jor}
\bibinfo{author}{\bibfnamefont{A.~M.} \bibnamefont{Green}} \bibnamefont{and}
  \bibinfo{author}{\bibfnamefont{B.~J.} \bibnamefont{Kavanagh}},
  \bibinfo{journal}{J. Phys. G} \textbf{\bibinfo{volume}{48}},
  \bibinfo{pages}{043001} (\bibinfo{year}{2021}), \eprint{2007.10722}.

\bibitem[{\citenamefont{Carr et~al.}(2021)\citenamefont{Carr, Kohri, Sendouda,
  and Yokoyama}}]{Carr:2020gox}
\bibinfo{author}{\bibfnamefont{B.}~\bibnamefont{Carr}},
  \bibinfo{author}{\bibfnamefont{K.}~\bibnamefont{Kohri}},
  \bibinfo{author}{\bibfnamefont{Y.}~\bibnamefont{Sendouda}}, \bibnamefont{and}
  \bibinfo{author}{\bibfnamefont{J.}~\bibnamefont{Yokoyama}},
  \bibinfo{journal}{Rept. Prog. Phys.} \textbf{\bibinfo{volume}{84}},
  \bibinfo{pages}{116902} (\bibinfo{year}{2021}), \eprint{2002.12778}.

\end{thebibliography}

\end{document}